\documentclass[12pt]{article}

\usepackage{multirow}
\usepackage{tabularx}
\usepackage{graphicx}
\usepackage{times}
\usepackage{url}

\newenvironment{sciabstract}{%
\begin{quote} \bf}
{\end{quote}}

\title{Stringent constraints on neutron-star radii from multimessenger observations and nuclear theory}

\author{Collin D. Capano$^{1,2*}$, Ingo Tews$^{3}$, Stephanie M. Brown$^{1,2}$, Ben Margalit$^{4,5,6}$, \\Soumi De$^{6,7}$, Sumit Kumar$^{1,2}$, Duncan A. Brown$^{6,7}$, Badri Krishnan$^{1,2}$, \\
Sanjay Reddy$^{8,9}$\\
\normalsize{$^1$ Albert-Einstein-Institut, Max-Planck-Institut f{\"u}r Gravitationsphysik,}\\ \normalsize{Callinstra{\ss}e 38, 30167 Hannover, Germany,}\\
\normalsize{$^2$ Leibniz Universit{\"a}t Hannover, 30167 Hannover, Germany,}\\
\normalsize{$^3$ Theoretical Division, Los Alamos National Laboratory, Los Alamos, NM 87545, USA,}\\
\normalsize{$^4$ Department of Astronomy and Theoretical Astrophysics Center,}\\ \normalsize{University of California, Berkeley, CA 94720, USA,}\\
\normalsize{$^5$ NASA Einstein Fellow}\\
\normalsize{$^6$ Kavli Institute for Theoretical Physics, University of California, Santa Barbara, CA 93106, USA,}\\
\normalsize{$^7$ Department of Physics, Syracuse University, Syracuse NY 13244, USA,}\\
\normalsize{$^8$ Institute for Nuclear Theory, University of Washington, Seattle, WA 98195-1550, USA,}\\
\normalsize{$^9$ JINA-CEE, Michigan State University, East Lansing, MI, 48823, USA}\\
}

\date{}

\begin{document} 


\baselineskip24pt


\maketitle 

\begin{sciabstract}

The properties of neutron stars are determined by the nature of the matter that they contain. These properties can be constrained by measurements of the star's size. We obtain stringent constraints on neutron-star radii by combining multimessenger observations of the binary neutron-star merger GW170817 with nuclear theory that best accounts for density-dependent uncertainties in the equation of state. We construct equations of state constrained by chiral effective field theory and marginalize over these using the gravitational-wave observations. Combining this with the electromagnetic observations of the merger remnant that imply the presence of a short-lived hyper-massive neutron star, we find that the radius of a $1.4\,\rm{M}_\odot$ neutron star is $R_{1.4\,\mathrm{M}_\odot} = 11.0^{+0.9}_{-0.6}~{\rm km}$ (90\% credible interval). Using this constraint, we show that neutron stars are unlikely to be disrupted in neutron-star black-hole mergers; subsequently, such events will not produce observable electromagnetic emission.
\end{sciabstract}

Neutron stars are arguably the most interesting astrophysical objects in the multimessenger era. The gravitational waves, electromagnetic radiation, and neutrinos produced by a variety of neutron-star phenomena carry information about the mysterious dense matter in their cores. The nature of this matter contains important information needed to understand phases of matter encountered in Quantum Chromodynamics---the fundamental theory of strong interactions. A measurement of the neutron-star radius or its compactness is critical both to interpret multimessenger observations of neutron stars and to determine the equation of state of dense matter~\cite{Lattimer:2000nx}. Until recently, measurement of neutron-star radii relied on X-ray observations of quiescent and accreting neutron stars. These analyses typically obtained radii in the range $10$--$14$~km and with poorly understood systematics~\cite{Ozel:2016oaf}, although this situation is likely to improve with recent observations by NICER~\cite{NICER1}. The multimessenger observation of the fortuitously close binary neutron-star merger GW170817~\cite{TheLIGOScientific:2017qsa} and its electromagnetic counterparts~\cite{GBM:2017lvd} provides information that can independently and more accurately determine neutron-star radii.

We combine state-of-the-art low-energy nuclear theory, constrained by experimental data, with multimessenger observations of the binary neutron-star merger GW170817~\cite{TheLIGOScientific:2017qsa} to measure the radii of neutron stars and to constrain the nuclear equation of state. Using conservative assumptions on the nuclear physics and the properties of the electromagnetic counterpart, we obtain the most stringent constraints on neutron-star radii to date. We find that the radius of a $1.4\,\mathrm{M}_\odot$ neutron star is $R_{1.4\,\mathrm{M}_\odot} = 11.0^{+0.9}_{-0.6}~{\rm km}$. 

Previous analyses of GW170817 have either neglected constraints on the equation of state from nuclear physics, or used a parameterization that was informed by a large number of nuclear physics models~\cite{TheLIGOScientific:2017qsa,De2018,Abbott:2018exr,Abbott:2018wiz,Radice&Dai18}. Neither approach properly accounts for the density-dependent theoretical uncertainties in our current understanding of dense matter.
We employ, for the first time, a strategy that allows us to overcome this deficiency. At low density, chiral effective field theory provides a description of matter in terms of nucleons and pions in which interactions are expanded in powers of momenta and include all operators consistent with the underlying symmetries of strong interactions~\cite{Weinberg1990,Weinberg1991,vanKolck:1994yi,Epelbaum:2008ga,Machleidt:2011zz}. This expansion defines a systematic order-by-order scheme that can be truncated at a given order and, most importantly, enables reliable theoretical uncertainty estimates from neglected contributions. Quantum Monte Carlo techniques are then used to solve the many-body Schr\"odinger equation to obtain the equation of state~\cite{Carlson:2015}. The resulting equation of state is characterized by an uncertainty which grows with density in a manner that can be justified from fundamental theory. We extend the equations of state to higher densities where the low-energy effective-field-theory expansion breaks down in a general way~\cite{Tews2018,Hebeler:2010jx}, while ensuring that the speed of sound is less than the speed of light and that the equations of state support a two-solar-mass neutron star~\cite{Antoniadis:2013pzd}.
The multimessenger observations of GW170817 are then used to constrain these equations of state to ensure that they are consistent with: (i) the detected gravitational waves during the inspiral; (ii) the production of a post-merger remnant that does not immediately collapse to a black hole (BH); and (iii) the constraints that the energetics of the gamma-ray burst and kilonova place on the maximum mass of neutron stars, $M_{\rm max}$. 

In neutron matter, chiral effective-field-theory interactions are expected to provide a good description of the equation of state up to $1-2~n_{\rm sat}$ where $n_{\rm sat}=0.16 \textrm{ fm}^{-3}$ is the nuclear saturation density. Earlier calculations show that uncertainties grow quickly with density, from roughly 30\% in the energy at saturation density to a factor of 2
at twice saturation density \cite{Tews2018}. To be conservative, we generate two collections of nuclear equations of state that differ in the density range restricted by \textit{ab initio} calculations: chiral effective-field-theory constraints are either enforced up to nuclear saturation density or up to twice nuclear saturation density. These two sets of equations of state are extended to higher densities as described earlier, and allow to study the impact of the assumption that chiral EFT remains valid up to twice nuclear
saturation density. 
For each of the two families, we generate 2000 individual equations of state distributed so that the radius of a $1.4\,\mathrm{M}_\odot$ neutron star is approximately uniform for each set. The resulting prior on $R_{1.4\,\mathrm{M}_\odot}$ is shown in the left-most panel of Figure 1. For each set of equations of state, we use stochastic samplers~\cite{Biwer:2018osg} to compute the posterior probability so that the tidal polarizability of GW170817 is consistent with a specific  equation of state. When generating model gravitational waveforms, the stochastic sampler randomly draws the neutron-star masses from a uniform distribution between $1$ and $2\,\mathrm{M}_\odot$, and then randomly draws a specific equation of state to compute the tidal polarizability of each star. 
The sky-position and luminosity distance of the source are fixed to those of the electromagnetic counterpart~\cite{Soares-Santos:2017lru,Cantiello:2018ffy}. This procedure allows us to directly constrain nuclear effective field theories from the gravitational-wave observations and to compute marginalized posterior probabilities for the star's radii using a model-independent non-parametric approach.

The result of constraining nuclear theory with the gravitational-wave observations is shown in the second panel of Figure 1. We find that the gravitational-wave observation GW170817 constrains the maximum radius of neutron stars, but is not informative at low radii, consistent with previous analyses~\cite{De2018,Abbott:2018exr}. The lower limit on the radius is set by nuclear theory and the requirement that the equation of state must support a neutron star of at least $1.9\,\mathrm{M}_\odot$~\cite{Antoniadis:2013pzd}. If one assumes that the chiral effective-field-theory description is valid only up to nuclear saturation density, it is possible to obtain large neutron stars that are not consistent with the tidal polarizability constraint from GW170817. In contrast, if a description in terms of nuclear degrees of freedom remains valid up to twice nuclear saturation density and the effective-field-theory approach can be applied, as suggested by earlier work~\cite{Tews:2018kmu}, then nuclear theory predicts neutron-star radii and tidal deformabilities that are consistent with GW170817. Simpler phenomenological models for the equation of state that are uninformed by nucleon-nucleon scattering data and which predict considerable stiffening of the equation of state between $n_{\rm sat}$ and $2n_{\rm sat}$ are excluded by GW170817.

Additional constraints on the equation of state can be obtained from the electromagnetic counterparts to GW170817. Since modeling the counterparts is challenging, we use a conservative approach that relies only on qualitative inferences from the kilonova and gamma-ray burst observations. The properties of these counterparts are inconsistent with either direct collapse to a black hole or the existence of a long-lived neutron-star remnant~\cite{Margalit&Metzger17,Bauswein+17}. 
This allows us to place two further constraints on the allowed equations of state. 

First, we discard samples from the posterior in which the total gravitational mass of the binary exceeds the threshold for prompt collapse to a black hole. Several approaches have been taken to calculate this threshold mass~\cite{Bauswein+13, Koppel:2019pys}. Here, for each equation of state in our sample we use relations calibrated to numerical relativity simulations of Ref.~\cite{Bauswein+13}, including uncertainties.
The effect of this constraint on the neutron-star radius is shown in the third panel of Figure 1, and significantly constrains the lower limit on $R_{1.4\,\mathrm{M}_\odot}$ \cite{Bauswein+17}. Second, we apply an upper limit on the maximum mass of neutron stars implied by the inconsistency of the electromagnetic counterparts with a long-lived neutron-star remnant~\cite{Margalit&Metzger17}.
We adopt a conservative estimate for this limit, $M_{\rm max} < 2.3\,\mathrm{M}_\odot$~\cite{Shibata:2019ctb}, consistent also with the 68.3\% credible interval of the recently reported $2.14^{+0.10}_{-0.09} \,\mathrm{M}_\odot$ pulsar~\cite{Cromartie:2019kug}. The result of applying this constraint is shown in the fourth panel of Figure 1. 

When constraining the allowed equations of state to those for which the maximum neutron-star mass is less than $2.3 \,\mathrm{M}_\odot$, we find that the predicted range for $R_{1.4\,\mathrm{M}_\odot}$ does not significantly change for any prior we investigated. This implies that there is no correlation between $R_{1.4\,\mathrm{M}_\odot}$ and $M_{\rm max}$. Such a correlation is typically found for smooth equations of state, e.g., equations of state that assume a description in terms of nucleons to be valid in the whole neutron star. In this case, limiting $M_{\rm max}$ would also constrain $R_{1.4\,\mathrm{M}_\odot}$. The general sets of equations of state we use here, however, include those with phase transitions that generally break this correlation and effectively decouple the high-density equation of state, which sets $M_{\rm max}$, from the low-density equation of state, that determines $R_{1.4\,\mathrm{M}_\odot}$~\cite{Tews2018}. As a consequence, we include equations of state with the largest possible $R_{1.4\,\mathrm{M}_\odot}$ but sufficiently small maximum masses, so that enforcing an upper limit on $M_{\rm max}$ has a negligible impact on the predicted radius range.
This decoupling highlights the importance of methods constraining the equation of state in different density regimes.

The right-most panel of Figure 1 compares our results to previous analyses~\cite{De2018,Abbott:2018exr}. Our constraint of $R_{1.4\,\mathrm{M}_\odot} = 11.0^{+0.9}_{-0.6}~{\rm km}$ is the most stringent bound on the neutron-star radius to date by a factor of $\simeq 2$. In Figure 2, we show the resulting mass-radius relation for our two equation-of-state sets and the marginalized posterior distributions of the component masses and radii for the two neutron stars in GW170817. In Table 1 we summarize our findings for the radii, tidal polarizabilities, and maximum neutron-star masses for the two equation-of-state sets. In addition, we present results for the maximum pressure explored in any neutron star, $P_{\rm max}$, and the pressure at four times nuclear saturation density, $P_{4n_{\rm sat}}$.

Comparing the constraints summarized in Table 1 for both equation-of-state sets, i.e., equations of state constrained by chiral effective field theory up to $n_{\rm sat}$ with those constrained up to $2n_{\rm sat}$, indicates that both effective-field-theory-based predictions for the equation of state are consistent with each other and with observations. These findings suggests that, in the absence of phase transitions in this density regime, the effective-field-theory description of neutron-rich matter remains useful and reliable up to $2n_{\rm sat}$ and excludes a considerable stiffening of the equation of state between $(1-2)~n_{\rm sat}$. Despite the larger uncertainties in the equation of state at higher densities, the electromagnetic and gravitational-wave observations can be combined with effective-field-theory-based equations of state up to $2 n_{\rm sat}$---for the first time---to greatly improve the constraint on the neutron-star radius. This has important implications for dense-matter physics and astrophysics. 

For dense-matter physics, we are able to derive robust constraints on the pressure of matter at moderate densities by combining our low density equations of state with the lower bound on the neutron-star radius derived from electromagnetic observations, and the upper bound from gravitational-wave observations. The pressure at $4 n_{\rm sat}$ is found to be $P_{\rm 4 n_{sat}}=161^{+58}_{-46}$ MeV/fm$^3$. This, taken together with the lower pressures predicted by nuclear theory in the interval $(1-2)~n_{\rm sat}$, supports earlier claims that the speed of sound in massive neutron stars must exceed $c/\sqrt{3}$~\cite{Bedaque:2014sqa}, where $c$ is the speed of light. We also provide improved estimates of the maximum pressure that can be realized inside neutron stars~\cite{LattimerPrakash2010}, $P_{\rm max}\leq 890$ MeV/fm$^3$. 

Our constraints on the neutron-star radius and polarizability impact the ability of gravitational-wave observations to distinguish between binary black-hole mergers and mergers containing neutron stars~\cite{Hannam:2013uu}. The gravitational-wave observations of GW170817 alone do not rule out the possibility that one or both objects in the merger were black holes~\cite{LIGOScientific:2019eut}. If Advanced LIGO and Virgo were to observe a source at comparable distance to GW170817 once they reach design sensitivity its signal-to-noise ratio would be $100$. Using simulated signals of this amplitude, we find that the gravitational waves could easily distinguish between a binary black-hole merger and the merger of two neutron stars governed by the SLy or AP4 equations of state with a Bayes factor greater than $10^6$.

Detecting the presence of matter from the inspiral of the compact objects is more challenging for neutron-star--black-hole binaries. We simulated a neutron-star--black-hole binary containing $10\,M_\odot$ black hole and a $1.4\, M_\odot$ neutron star that has a dimensionless tidal polarizability at the upper bounds of our $90\%$ credible interval, $\Lambda = 370$. We place the source at the same distance as GW170817 and assume that the detectors are at design sensitivity~\cite{LIGO:aligodesign}; this binary has a signal-to-noise ratio of $190$. We calculate the Bayes factor comparing a neutron-star--black-hole model to a binary-black-hole model (i.e. zero tidal deformability for both compact objects). Even at this extremely large signal-to-noise ratio, we find that the Bayes factor is $\sim 1$ meaning that the models are indistinguishable. The inspiral waveforms of binary black holes and neutron-star--black-hole mergers become less distinguishable as the neutron-star mass increase, the polarizability decreases, or the black-hole mass increases, and any effect of matter becomes even harder to measure with gravitational waves alone. Electromagnetic counterparts or post-merger signatures will therefore be critical to distinguish between binary black-hole, binary neutron star, and neutron-star--black-hole mergers observed by Advanced LIGO and Virgo \cite{Hinderer:2018pei}.

The composition and amount of ejecta from binary neutron stars and neutron-star--black-hole binaries, which powers the electromagnetic emission, is sensitive to the neutron-star radius \cite{Foucart12,Bauswein+13a,Hotokezaka+13}. Our limits have implications for electromagnetic signatures and their observability. This is especially true for mergers such as S190814bv, which was recently reported by the LIGO and Virgo collaborations~\cite{GCN25333}. A kilonova or gamma-ray burst counterpart is only expected if the neutron star is tidally disrupted before the merger; a condition which depends crucially on the neutron-star radius. Figure 3 summarizes the parameter space of neutron-star--black-hole mergers where a mass ejection (and a corresponding electromagnetic counterpart) is expected
based on fits to numerical relativity simulations \cite{Foucart+18}.
Our novel constraints on $R_{1.4\,\mathrm{M}_\odot}$ imply that 
$\sim 1.4\,\mathrm{M}_\odot$ neutron stars cannot be disrupted in such mergers by non-spinning black holes,
unless the black-hole mass is unusually low ($<3.4\,\mathrm{M}_\odot$).
More generally, our constraints on neutron-star radii will be useful to predict and test correlations between electromagnetic and gravitational-wave observations in the future~\cite{Margalit&Metzger19}. 

Our improved constraints on the neutron-star radius have implications for the interpretation of electromagnetic observations of neutron stars, in particular for X-ray observations of accreting neutron stars in low mass X-ray binaries (LMXBs).  Recent observations suggest the presence of accretion-driven heating and cooling of the neutron-star inner crust~\cite{Wijnands:2017jsc}. The interpretation of the X-ray light curves in terms of the fundamental properties of the matter in the inner crust such as its thermal conductivity and specific heat is sensitive to the assumed neutron-star radius \cite{Brown:2009kw}.  Similarly, the radius is a key parameter in models of X-ray bursts and quiescent surface emission of neutron stars in LMXBs. In the past, this sensitivity has been exploited to place constraints on neutron-star radii. Knowing the radius could shed light on other astrophysical aspects of these commonly observed X-ray phenomena \cite{2006csxs.book}.

Observations of neutron star mergers with higher signal-to-noise ratios will improve these constraints. Using a simulated signal at a signal-to-noise ratio of $\sim100$, we find that the gravitational-wave signal alone would improve the constraint on $R_{1.4\,\mathrm{M}_\odot}$ by a factor of 2 relative to our result that uses the multi-messenger observations of GW170817. For a signal of this amplitude, we will be able to place a lower bound on the neutron star radius from the gravitational-wave observation, independent of the electromagnetic counterpart. However, to realize this measurement in practice, improved waveforms will be needed so that modelling systematics do not bias the measurement\cite{Barkett:2015wia,Narikawa:2019xng}.

Over the next ten years, the LIGO and Virgo detectors are expected to gain an additional factor of $\sim3$ in sensitivity \cite{Aasi:2013wya}. In addition, the Japanese Kagra detector and LIGO India are expected to be operational. With these improvements, we estimate the rate of binary neutron star mergers with a signal-to-noise ratio $\geq 100$ to be one per $4^{+36}_{-3}$ years~\cite{LIGOScientific:2018mvr}. Constraints on $R_{1.4\,\mathrm{M}_\odot}$ from gravitational-wave observations alone will be dominated by these relatively rare, high signal-to-noise ratio events. In contrast, multimessenger methods of constraining the equation of state can provide significant improvements even for low signal-to-noise events with detected electromagnetic counterparts \cite{Margalit&Metzger19}. These methods are thus extremely promising, however their susceptibility to systematics will need to be better understood.

We combined multimessenger observations of the binary neutron-star merger GW170817 and the best current knowledge of the uncertainties associated with the equation of state of dense matter to determine the neutron-star radius. This also allows us to place stringent bounds on the pressure of matter at moderate density where theoretical calculations remain highly uncertain. Our robust upper bound on the neutron-star radius of $R_{1.4\,\mathrm{M}_\odot} = 11.0^{+0.9}_{-0.6}~{\rm km}$ is a significant improvement, with important implications for multimessenger astronomy and nuclear physics.  To allow our results to be used by the community for further analysis we provide the equations of state used as our prior and the full posterior samples from our analysis as supplemental materials.

\section*{Methods}

\subsection*{Nuclear equations of state from chiral effective field theory}

Nuclear effective-field-theory methods~\cite{Epelbaum:2008ga,Machleidt:2011zz} represent a consistent and efficient way of constructing models for nuclear interactions while incorporating symmetries of the fundamental theory for strong interactions, as well as low-energy constraints from nuclear experiments. This is especially useful when extrapolating nuclear interactions to regimes where experimental data is scarce or not available, in particular for neutron-rich systems.  Among nuclear effective field theories, chiral effective field theory starts from the most general Lagrangian containing both pions and nucleons, consistent with all the fundamental symmetries for nuclear interactions. Since this Lagrangian has an infinite number of terms, the separation of scales between typical momenta in nuclear systems and all heavier degrees of freedom is used to expand the Lagrangian in powers of $p/\Lambda_b$. 
Here, $p$ is the typical momentum scale of the system at hand $\Lambda_b\approx 600\,$Mev~\cite{Melendez:2017phj} is the breakdown scale, which determines when heavier degrees of freedom become important and the effective field theory breaks down. This expansion defines a systematic order-by-order scheme for the interaction, that can be truncated at a given order and enables the estimation of reliable theoretical uncertainties from neglected contributions. Chiral effective field theory describes nuclear interactions in terms of explicitly included long-range pion-exchange interactions and parameterized short-range contact interactions. These short-range interactions depend on a set of unknown low-energy couplings, that absorb all unresolved high-energy degrees of freedom, and are adjusted to reproduce experimental data. 

To generate our families of equations of state, we start from microscopic quantum Monte Carlo calculations of the neutron-matter equation of state using two different nuclear Hamiltonians from chiral effective field theory up to $2n_{\rm sat}$~\cite{Tews:2018kmu}. The two Hamiltonians used in this work were fit to two-nucleon scattering data, the binding energy of the alpha particle, and the properties of neutron-alpha scattering, and reliably describe these systems~\cite{Lynn:2015jua}. They also have been benchmarked in calculations of nuclei up to $^{16}$O with great success~\cite{Lonardoni:2017hgs,Lynn:2019rdt}. In neutron matter, the limit of applicability of these effective-field-theory interactions has been estimated to be around twice nuclear saturation density~\cite{Tews:2018kmu}. Each of the two Hamiltonians we employ here has an associated theoretical uncertainty band stemming from the above-mentioned truncation of the chiral series at a finite order. These bands serve as an estimate for the uncertainty due to the limited description of nuclear interactions. The difference between the two Hamiltonians explores the remaining scheme and scale dependence of the chiral interactions. These two sources of uncertainty dominate the neutron-matter calculations.

From these neutron-matter calculations, we then construct the neutron-star equation of state by extending the pure neutron matter results to beta equilibrium and adding a crust~\cite{Tews:2016ofv}. This allows us to extend our neutron-matter uncertainties to the equation of state of neutron stars up to $2n_{\rm sat}$. At higher densities, chiral effective field theory breaks down because short-range details that are not resolved in the chiral effective-field-theory description become important. To be able to describe neutron stars up to the highest masses, we need to extend our calculations to higher densities in a general and unbiased fashion, i.e., without making assumptions about the properties of the equation of state or its degrees of freedom. To achieve this, we use the results from our microscopic calculations where they remain reliable, up to a density $n_{\rm tr}$ which we choose to be either $n_{\rm sat}$ or $2n_{\rm sat}$. We then compute the resulting speed of sound, $c_S$, in neutron-star matter with its uncertainty band for the two Hamiltonians. For each Hamiltonian, we select a $c_S$ curve up to $n_{\rm tr}$ from the uncertainty band, by sampling a factor $f_{\rm err}\in [-1,1]$ which interpolates between upper and lower uncertainty bound. At densities above $n_{\rm tr}$, we sample a set of points $c_S^2(n)$ randomly distributed between $n_{\rm tr}$ and $12 n_{\rm sat}$, and connect these points by line segments. We only require the speed of sound to be positive and smaller than the speed of light, $c$, i.e., the resulting curve has to be stable and causal. For each such curve, we construct a related curve that includes a strong first-order phase transition, by replacing a segment with a random onset density and width with a segment with $c_S=0$. We then reconstruct the equation of state from the resulting curve in the $c_S$ plane and solve the Tolman-Oppenheimer-Volkoff equations. We have explored the sensitivity of neutron-star properties to the number of points in the $c_S$ plane, and constructed extensions with 5-10 points. The differences between these different extensions have been found to be very small. For the equations of state explored here, we chose a 6-point extension. We repeat this procedure for equal numbers of equations of state of $\mathcal{O}(10,000)$ for the two microscopic Hamiltonians~\cite{Tews:2019cap}.

The resulting family of equations of state is constrained by low-energy nuclear theory as well as general considerations on stability and causality. Finally, we enforce that each equation of state reproduces a neutron star with at least $1.9 M_{\odot}$, which is a conservative estimate for the lower uncertainty bound for the two-solar-mass neutron-star observations~\cite{Antoniadis:2013pzd}. For each of the two families of equations of state, one for $n_{\rm tr}=n_{\rm sat}$ and one for $n_{\rm tr}=2 n_{\rm sat}$, we then randomly select 2000 equations of state that have a uniform prior in $R_{1.4}$.

\subsection*{Gravitational wave parameter estimation}

We use Bayesian methods to measure the tidal polarizability of GW170817 and to infer the equations of state that are most consistent with the observations.  Given time-series data from the Hanford, Livingston, and Virgo detectors $\vec{d} = \left\{\vec{d}_H, \vec{d}_L, \vec{d}_V\right\}$ and a model waveform $h$, the probability that the binary has a set of parameter values $\vec{\vartheta}$ is
\begin{equation}
\label{eqn:bayes_theorem}
    p(\vec{\vartheta}|\vec{d}, h; I) = \frac{p(\vec{d}|\vec{\vartheta}, h; I)p(\vec{\vartheta}|h; I)}{p(\vec{d}|h; I)},
\end{equation}
where $p(\vec{d}|\vec{\vartheta}, h; I)$, $p(\vec{\vartheta}|h; I)$, and $p(\vec{d}|h; I)$ are the likelihood, prior, and evidence, respectively. The $I$ indicates additional assumed information, such as the field theory used to describe nuclear interactions. We assume that the detector noise is wide-sense stationary colored Gaussian noise with zero mean, and is independent between observatories. In that case, the likelihood is
\begin{equation}
\label{eqn:likelihood}
    p(\vec{d} | \vec{\vartheta}, h; I) \propto \exp\left[-\frac{1}{2} \sum_{i=H,L,V} \left<\vec{d}_i - \vec{h}_i(\vec{\vartheta}), \, \vec{d}_i - \vec{h}_i(\vec{\vartheta})\right>\right],
\end{equation}
where the brackets $\left<\cdot,\cdot\right>$ indicate an inner product that is weighted by the inverse power spectral density of the noise in each detector.

We use the gravitational-wave data associated with the GWTC-1 release~\cite{LIGOScientific:2018mvr} from the GW Open Science Center (GWOSC)~\cite{Vallisneri:2014vxa}. Specifically we use the 4096~s duration $16\,384$~Hz sampled frame data for GW170817 from the list of GWTC-1 confident detections, which we downsample to $4096\,$Hz. These data contain a non-Gaussian noise transient in the L1 data, which we remove by subtracting the glitch model made available in LIGO document LIGO-T1700406. We include these glitch-subtracted data in our data release. Two hundred seconds of data spanning $[t_0-190\,\mathrm{s}, t_0+10\,\mathrm{s})$ are filtered starting from $20\,$Hz, where $t_0=1187008882.443$ is an estimate of the geocentric GPS time of the merger obtained from the modeled searches that detected GW170817 \cite{TheLIGOScientific:2017qsa}. The power spectral density of the noise is estimated using a variant of Welch's method \cite{Allen:2005fk} on $1632\,$s of data that precedes the start of the analysis time.

To sample the posterior probability over the full parameter space we use Markov-chain Monte Carlo (MCMC) \cite{ForemanMackey:2012ig, Vousden:2015} and Nested Sampling stochastic samplers \cite{dynesty} in the PyCBC Inference framework \cite{Biwer:2018osg}. The resulting probability-density function can be numerically marginalized to provide estimates of single parameters. Marginalizing $p(\vec{d}|\vec{\vartheta}, h; I)p(\vec{\vartheta}|h; I)$ over all parameters provides an estimate of the evidence $p(\vec{d}|h; I)$. Taking the ratio of evidences for different physical models $p(\vec{d}|h; B)/p(\vec{d}|h;A)$ provides the Bayes factor $\mathcal{B}_{A}(B)$, which quantifies how much the data supports model $B$ relative to model $A$. We assume that the binary consists of two compact objects with spins aligned with the orbital angular momentum, and that the binary has negligible eccentricity by the time it can be detected by the LIGO and Virgo interferometers. Under these assumptions, the observed gravitational wave depends on 13 parameters: six ``intrinsic" parameters -- the mass $m_{1,2}$, dimensionless-spin magnitude $\chi_{1,2}$, and tidal polarizability $\Lambda_{1,2}$ of each component star -- and seven ``extrinsic" parameters -- the binary's right ascension $\alpha$, declination $\delta$, luminosity distance $d_L$, inclination $\iota$, coalescence time $t_c$, reference phase $\phi$, and polarization $\psi$. We fix the sky location and luminosity distance in our analysis to $\alpha = 13^{\mathrm{h}}\, 09^{\mathrm{m}}\, 48.1^{\mathrm{s}}$, $\delta = -23^{\circ}\, 22'\, 53.4"$ \cite{Soares-Santos:2017lru}; $d_L = 40.7\,\mathrm{Mpc}$ \cite{Cantiello:2018ffy}. The phase $\phi$ is analytically marginalized over using a prior uniform between $0$ and $2\pi$. We also use uniform priors on $\psi \in [0, 2\pi)$, $\cos \iota \in [-1, 1)$, and $t_c \in t_0 \pm 0.1\,$s. For the dimensionless spin components $\chi_{1,2}$ we use a prior uniform in $[-0.05, 0.05)$. This is consistent with the fastest-known pulsar in a double neutron-star system \cite{Burgay:2003jj}, and was used in previous studies of GW170817 \cite{De2018, Abbott:2018exr, TheLIGOScientific:2017qsa}.

Observations of millisecond pulsars yield a large variance in possible neutron-star masses, with the largest observed masses estimated to be $2.01 \pm 0.04\,\mathrm{M}_\odot$ \cite{Burgay:2003jj} and $2.17^{+0.11}_{-0.10} M_\odot$ \cite{Cromartie:2019kug}. 
We therefore use a prior  distribution uniform in $[1, 2)~\mathrm{M}_\odot$ for the detector frame component masses in our main analysis.  Assuming the standard $\Lambda$-CDM cosmology \cite{Ade:2015xua}, this corresponds to $m_{1,2}/\mathrm{M}_\odot \sim U(0.99, 1.98)$ in the source frame at 40.7~Mpc.
Electromagnetic observations of double neutron-star systems in the galaxy have yielded a best-fit neutron-star mass distribution of $p(m_{\mathrm{NS}}/\mathrm{M}_\odot) \sim \mathcal{N}(\mu = 1.33, \sigma = 0.09)$ \cite{Ozel:2016oaf}. We repeated our analysis using this distribution as our prior on each component mass and we find that our results are insensitive to the choice of mass prior.

We directly sample over individual equations of state instead of $\Lambda_{1,2}$ or $\Lambda_{s}$. For each of the 2000 equations of state for each model, we order the equations by the radius they yield for a $1.4\,\mathrm{M}_\odot$ neutron star, $R_{1.4\,\mathrm{M}_\odot}$. The equations of state are generated such that the distribution of $R_{1.4\,\mathrm{M}_\odot}$ is approximately uniform in the range supported. This results in the marginal prior on each star's radius $R_{1,2}$ to also be approximately uniform for both of our mass priors, since the radii do not vary much over the mass ranges considered. We then sample over an equation-of-state index $k_{\mathrm{EOS}} \sim U[1,2000]$. Using the index and the two component masses $m_{1,2}$, we calculate $\Lambda_{1,2}$, with which we generate a model gravitational wave $h(\vec{\vartheta})$ and measure the likelihood, Eq.~(\ref{eqn:likelihood}). In this manner, we ensure that both component masses use exactly the same equation of state, with all sampled equations of state being constrained by chiral effective field theory.

We use restricted TaylorF2 post-Newtonian waveforms~\cite{Sathyaprakash:1991mt,Buonanno:2009zt,Mikoczi:2005dn,Arun:2008kb,Bohe:2013cla,Vines:2011ud} in our analysis. The effect of analyzing GW170817 using different waveforms was studied in Ref.~\cite{Abbott:2018wiz} with unconstrained equations of state. To test the effect of using TaylorF2 waveforms in this study, we repeat the $n_{\mathrm{sat}}$ analysis with the uniform mass prior using PhenomDNRT~\cite{PhysRevD.96.121501,PhysRevD.99.024029,PhysRevD.93.044006,PhysRevD.93.044007}. We also tested whether increasing the sample rate to $8192\,$Hz had any effect using this waveform model. We found negligible differences in all three cases; we therefore only report results using the TaylorF2 model.

\subsection*{Constraints on neutron star radii from multimessenger observations}

Observations of the kilonova associated with GW170817 indicate that a large $\sim 0.02-0.08 \,\mathrm{M}_\odot$ amount of mass was ejected, and that this ejecta must contain components with both large and small electron fraction~\cite{Villar+17}. 
These inferences are inconsistent with numerical simulations of binary neutron-star mergers in which the remnant promptly (within milliseconds) collapses to form a black hole. These simulations generally find a low amount of ejected matter with only high electron fraction, at odds with the optical and near-infrared observations of GW170817.
Therefore, electromagnetic observations of GW170817 are inconsistent with a prompt collapse to a black-hole.

Systematic numerical studies of binary neutron-star mergers have found that the condition for prompt collapse depends primarily on the total binary mass in comparison to an equation-of-state-dependent threshold mass \cite{Bauswein+13,Bauswein+17a,Koppel:2019pys}. (See \cite{Kiuchi+19} for possible effects of large mass ratios; however, such mass ratios are not expected if neutron stars are drawn from a distribution similar to the population of Galactic binary neutron stars~\cite{Margalit&Metzger19}.) Bauswein et al.~\cite{Bauswein+13,Bauswein+17a} show that this threshold mass increases as a function of the neutron-star radius and maximum mass, following
\begin{equation}
    \label{eq:threshold_mass}
    M_{\rm thresh} \approx M_{\rm max} \left( 2.380 - 3.606 \frac{G M_{\rm max}}{c^2 R_{1.6\,\mathrm{M}_\odot}} \right) \pm 0.05 \,\mathrm{M}_\odot .
\end{equation}
We apply this condition to posterior samples for each equation of state in our analysis to impose the requirement that the binary should not promptly collapse into a black hole upon merger. For each equation of state in our sample we calculate the threshold mass using the above expression. To account for the systematic error in Eq.~\ref{eq:threshold_mass}, a random draw from a normal distribution with a standard deviation of 0.05 is added to each threshold mass sample. We then discard samples from the posterior for which $M_{\rm total} < M_{\rm thresh}$. This rules out equations of state with low $R_{1.4\,\mathrm{M}_\odot}$, as shown in Figure 1.
Bauswein et al.~\cite{Bauswein+17} first used similar methods to place a lower bound of $R_{1.6 \,\mathrm{M}_\odot} \geq 10.64 {\rm km}$. However, that study required equation-of-state independent assumptions regarding causality to relate $M_{\rm max}$ to $R_{1.6\,\mathrm{M}_\odot}$.
Additionally, they imposed a more stringent constraint $M_{\rm total} < M_{\rm thresh} - 0.1M_\odot$ to obtain this value, and find instead $R_{1.6\,\mathrm{M}_\odot} > 10.27 \, {\rm km}$ for the more conservative assumption $M_{\rm total} < M_{\rm thresh}$ that we adopt here. Our results are consistent with these previous findings, but manage to place a slightly stronger lower limit on the radius (in the conservative case). 

The electromagnetic observations of GW170817 are also inconsistent with a long-lived merger remnant. If even a small fraction of the remnant's rotational energy is extracted through electromagnetic torques (as expected if the remnant neutron star develops even a modest external dipole magnetic field) this would deposit sufficient energy into the surrounding medium to be incompatible with energetic constraints from the kilonova and gamma-ray burst afterglow modelling~\cite{Margalit&Metzger17}. A long-lived neutron star would also be in tension with the observed $1.7$s delay between the gamma-ray burst and merger~\cite{Shibata+17,Rezzolla+18,Ruiz+18}.
This requirement places an upper limit on $M_{\rm max}$ of roughly $\sim 2.2\,\mathrm{M}_\odot$~\cite{Margalit&Metzger17,Shibata+17,Rezzolla+18,Ruiz+18}. To err on the side of caution we here adopt a more conservative estimate $M_{\rm max} < 2.3 \,\mathrm{M}_\odot$~\cite{Shibata:2019ctb}.
We implement this constraint by discarding samples whose equations of state do not satisfy this requirement on $M_{\rm max}$. As described in the main text, this has little affect on $R_{1.4\,\mathrm{M}_\odot}$ because our equations of state allow for the most general behavior above $n_{\rm sat}$ or $2 n_{\rm sat}$ including phase transitions, such that the high-density equation of state, which sets $M_{\rm max}$, is effectively decoupled from the low-density region that determines $R_{1.4\,\mathrm{M}_\odot}$.

\subsection*{Conditions for neutron-star tidal disruption}

Neutron-star--black-hole mergers have been studied extensively in the literature starting from the pioneering work of Lattimer and Schramm~\cite{LattimerSchramm74}, however an electromagnetic counterpart to such mergers has not yet been unambiguously detected.
The most promising counterparts, an optical/near-infrared kilonova or a gamma-ray burst, depend on whether significant matter can be stripped off the neutron star prior to merger.
The condition for the neutron star to be tidally disrupted before merger depends sensitively on the neutron-star radius, in addition to the neutron-star mass and black-hole mass and spin~\cite{Shibata+08,Foucart12,Kyutoku+15}.
Previous work has investigated this parameter space identifying regions where electromagnetic counterparts may be expected and their observational signatures (e.g. \cite{Foucart12,PannaraleOhme14,Barbieri+19}), 
however the unknown neutron-star radius introduced an inherent uncertainty in such analyses.
Our new constraint on $R_{1.4\,\mathrm{M}_\odot}$ allows us to reduce the uncertainty in this four-parameter space and provide more precise predictions on whether electromagnetic counterparts may be expected for neutron-star--black-hole mergers given $M_{\rm BH}$, $M_{\rm ns}$, and $\chi_{\rm BH}$ inferred from the gravitational-wave signal.

Foucart et al. \cite{Foucart+18} presented a systematic numerical study of mass ejection from neutron-star--black-hole mergers and provided a fitting formula for the amount of mass remaining outside the black-hole horizon shortly after merger and which could produce detectable electromagnetic emission, $M_{\rm det}$,
\begin{equation}
    M_{\rm det} \approx M_{\rm ns} \left[ \alpha \eta^{-1/3} \left( 1 - 2 \frac{GM_{\rm ns}}{c^2R_{\rm ns}} \right) - \beta \eta^{-1} \frac{R_{\rm ISCO}}{R_{\rm ns}} + \gamma \right]^\delta .
\end{equation}
In the above, $(\alpha,\beta,\gamma,\delta) = (0.406,0.139,0.255,1.761)$ are parameters fit to the numerical relativity simulations, $\eta$ is the symmetric mass ratio, and $R_{\rm ISCO}(\chi_{\rm BH})$ is the radius of the innermost stable circular orbit (ISCO) of the black hole and depends on its spin parameter $\chi_{\rm BH}$ \cite{Bardeen+72}.
In Figure 3 we have shown curves along which $M_{\rm det}=0$ as a function of the black-hole mass and spin, and for different neutron-star masses. Above each curve (higher spin) the neutron star would cross the black-hole ISCO before being tidally stripped of any matter, and a kilonova or gamma-ray burst counterpart would not be expected.

\subsection*{Prospects for tighter constraints on neutron-star radii}

We explore the prospects for improving constraints on the neutron-star radius and tidal deformability. We study the impact of a louder signal-to-noise ratio, and the choice of waveform models used in the likelihood computation for such loud signals. We generate realizations of stationary Gaussian noise for the Advanced LIGO and Virgo detectors, colored by the power-spectral densities representative of the design sensitivity of the detectors~\cite{LIGO:aligodesign}. A simulated signal with parameters representative of those for GW170817 is added to the noise. The equation of state determining the structure of this system is the median from our $2n_{\mathrm{sat}}$ analysis of GW170817. We place the source at a sky location such that an optimal contribution to the signal-to-noise ratio is obtained from the network of detectors. Specifically, the parameters of our simulated signal are $m_1 = 1.48\,\rm{M}_\odot$, $m_2 = 1.26\,\rm{M}_\odot$, $\chi_1 = -0.030$, $\chi_2 = -0.026$, $\Lambda_1 = 136$, $\Lambda_2 = 345$, $d_L = 40.7\,$Mpc, $\iota = 149^\circ$, $\psi = 273.8^\circ$, $t_c = 1187008882.4283648$, $\alpha = 16^{\mathrm{h}}\, 15^{\mathrm{m}}\, 4.9^{\mathrm{s}}$, and $\delta = -32^{\circ}\, 52'\, 5.16”$. We use the PhenomDNRT model to construct the simulated signal. The resulting injected signal has a signal-to-noise ratio of $\sim 100$. This represents the best possible scenario of observing GW170817 with the Advanced LIGO-Virgo detectors (keeping the luminosity distance unmodified). However, within the next decade the sensitivity of the LIGO detectors is expected to surpass the sensitivity we have assumed here~\cite{Aasi:2013wya}. It is therefore reasonable to expect that a binary neutron star will be detected at this signal-to-noise ratio within the next decade.

We perform a parameter estimation analysis on the simulated data using the same prior and settings as in the $2n_{\mathrm{sat}}$ analysis described above. To study the effect of waveform systematics, which are significant at these signal-to-noise ratios, we do an analysis using the TaylorF2 waveform model, and compare to an analysis using the PhenomDNRT model. Supplementary Figure 1 shows a comparison of the $R_{1.4\,\mathrm{M}_\odot}$ and $\tilde \Lambda$ posterior probability densities obtained from the two analyses of the simulated data with that obtained from our gravitational-wave only analysis of GW170817. Using TaylorF2, the measurements of $R_{1.4\,\mathrm{M}_\odot}$ and $\tilde \Lambda$ improve by factors of 1.6 and 1.8 respectively, compared to GW170817. With PhenomDNRT, the measurements of $R_{1.4\,\mathrm{M}_\odot}$ and $\tilde \Lambda$ are improved by factors of 2.9 and 3.2 relative to GW170817, respectively. This illustrates that at higher signal-to-noise ratio, parameter measurement accuracy is significantly improved with the use of better waveform models. 

\section*{Data availability}
All data is available in the manuscript or the supplementary materials. Full posterior data samples are available at \url{https://github.com/sugwg/gw170817-eft-eos}. The gravitational-wave data used in this work was obtained from the Gravitational Wave Open Science Center (GWOSC) at {https://www.gw-openscience.org}. 

\section*{Code availability}
All software used in this analysis is open source and available from \url{https://github.com/gwastro/pycbc}.

\section*{Acknowledgments}
We thank Bruce Allen, Wolfgang Kastaun, James Lattimer, and Brian Metzger for valuable discussions. \textbf{Funding:} This work was supported by U.S. National Science Foundation grants
PHY-1430152 to the JINA Center for the Evolution of the Elements (SR),
PHY-1707954 (DAB, SD);
U.S. Department of Energy grant DE-FG02-00ER41132 (SR);
NASA Hubble Fellowship grant \#HST-HF2-51412.001-A awarded by the Space Telescope Science Institute, which is operated by the Association of Universities for Research in Astronomy, Inc., for NASA, under contract NAS5-26555 (BM); and the U.S. Department of Energy, Office of Science, Office of Nuclear Physics, under Contract DE-AC52-06NA25396, the Los Alamos National Laboratory (LANL) LDRD program, and the NUCLEI SciDAC program (IT). 
DAB, SD, and BM thank the Kavli Institute for Theoretical Physics (KITP) where portions of this work were completed. KITP is supported in part by the National Science Foundation under Grant No. NSF PHY-1748958. 
Computational resources have been provided by Los Alamos Open Supercomputing via the Institutional Computing (IC) program, by the National Energy Research Scientific Computing Center (NERSC), by the J\"ulich Supercomputing Center, by the ATLAS Cluster at the Albert Einstein Institute in Hannover, and by Syracuse University. 
GWOSC is a service of LIGO Laboratory, the LIGO Scientific Collaboration and the Virgo Collaboration. LIGO is funded by the National Science Foundation. Virgo is funded by the French Centre National de Recherche Scientifique (CNRS), the Italian Istituto Nazionale della Fisica Nucleare (INFN) and the Dutch Nikhef, with contributions by Polish and Hungarian institutes.

\textbf{Authors contributions:}
Conceptualization, DAB, CDC, BK, BM, SR, IT;
Data curation, DAB, CDC, SD, IT;
Formal analysis, CDC, SMB, IT, SD;
Funding acquisition: DAB, BK, BM, SR, IT;
Methodology: DAB, CDC, SD, BK, BM, SR, IT;
Project administration: DAB, BK, SR, IT;
Resources: DAB, BK, IT;
Software: DAB, SMB, CDC, SD, SK, BM, IT;
Supervision: DAB, BK, SR;
Validation: DAB, SMB, CDC, SD, IT;
Visualization: SMB, CDC, BM;
Writing--original draft: DAB, SMB, CDC, IT;
Writing--review and editing: DAB, SMB, CDC, SD, BK, BM, SR, IT.
\textbf{Competing interests:} The authors declare no competing interests.
\clearpage

\begin{figure}[ht]
\centering
\includegraphics[width=\textwidth]{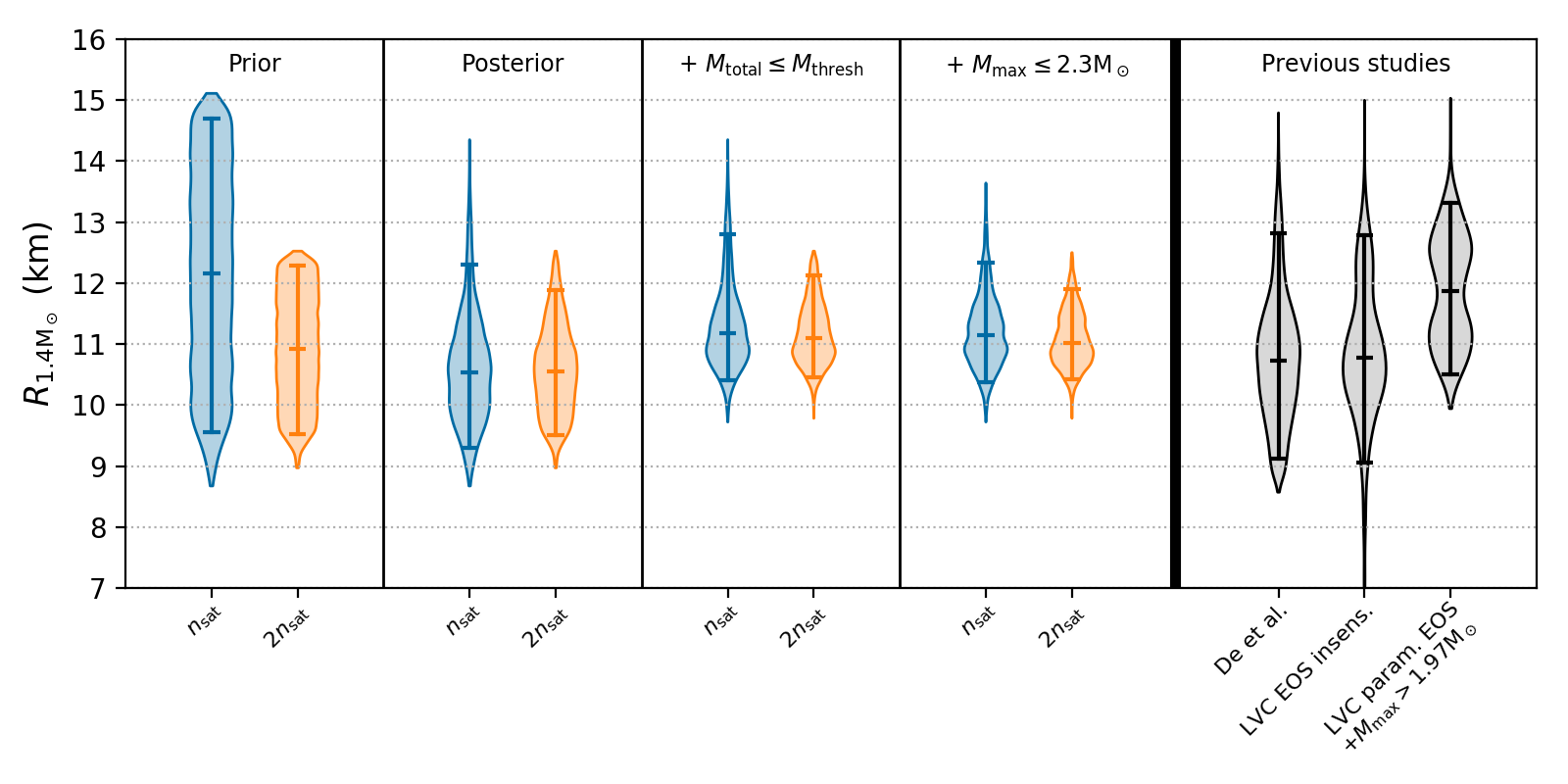}
\caption{Comparison of the estimated radius of a $1.4\,\mathrm{M}_\odot$ neutron star, $R_{1.4\,\mathrm{M}_\odot}$, at different stages of our analysis. In all panels, 1D marginal distributions are indicated by the shaded regions, with the median and the plus/minus 95th and 5th percentiles indicated by the lines. The left panel shows the marginalized prior on $R_{1.4\,\mathrm{M}_\odot}$, assuming chiral effective field theory up to $n_{\mathrm{sat}}$ (blue) and $2n_{\mathrm{sat}}$ (orange). Subsequent panels show the posterior on $R_{1.4\,\mathrm{M}_\odot}$ from the gravitational-wave analysis alone, the posterior with the constraint that the estimated total mass $M_{\mathrm{total}}$ to be less than the threshold mass for prompt collapse $M_{\mathrm{thresh}}$, and the posterior with the additional constraint that the maximum NS mass $M_{\rm max}$ supported by all equations of state $\leq 2.3\,\mathrm{M}_\odot$. The right panel shows posteriors on $R_{1.4\,\mathrm{M}_\odot}$ from De et al.~\cite{De2018}, and $R_1 \approx R_{1.4\,\mathrm{M}_\odot}$ from the equation-of-state-insensitive and parameterized equation-of-state analyses reported by Abbott et al.~\cite{Abbott:2018exr} (labelled LVC). In all analyses, a uniform prior was used on the component masses.
\label{fig:mainresults}}
\end{figure}

\clearpage

\begin{figure}
\centering
\includegraphics[width=\textwidth]{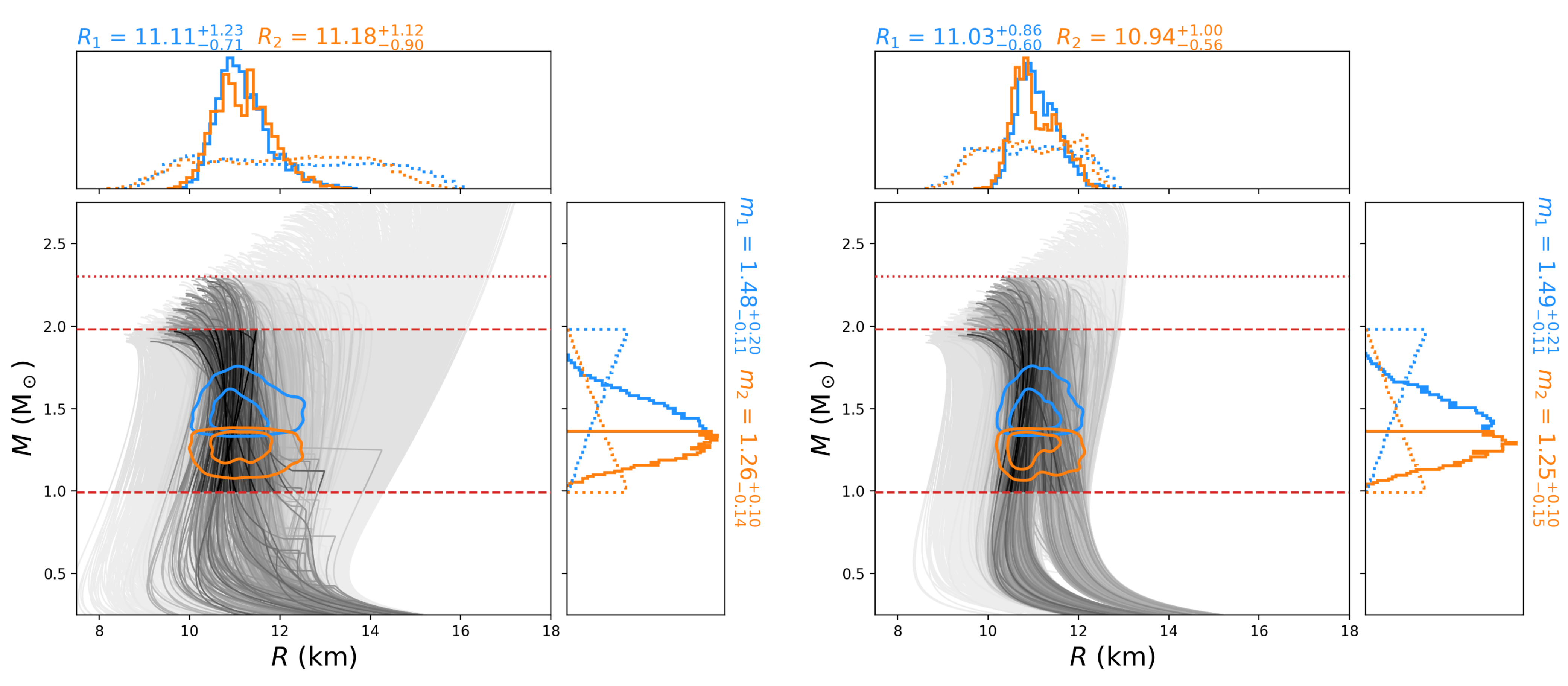}
\caption{Neutron-star mass-radius curves and marginalized posterior distributions of the source component masses $m_{1,2}$ and radii $R_{1,2}$, assuming a prior uniform in component masses, with chiral effective field theory enforced up to $n_{\mathrm{sat}}$ (left) and $2n_{\mathrm{sat}}$ (right) and all additional observational constraints enforced. The dashed, horizontal red lines indicate the range of masses spanned by the prior. The top dotted red line indicates the maximum neutron-star mass constraint. Any equation of state that has support above that line is excised. Each gray-black line represents a single equation of state, which we sample directly in our analysis. The shading of the lines is proportional to the marginalized posterior probability of the equation of state; the darker the line, the more probable it is. The contours show the 50th and 90th percentile credible regions (blue for the more massive component, orange for the lighter component). The 1D marginal posteriors are shown in the top and side panels; the corresponding priors (without electromagnetic constraints) are represented by the dotted blue and orange lines. Quoted values are the median plus/minus 95th and 5th percentiles. 
\label{fig:mr_posterior_with_constraint}}
\end{figure}

\clearpage

\begin{figure}[ht]
\centering
\includegraphics[width=0.9\textwidth]{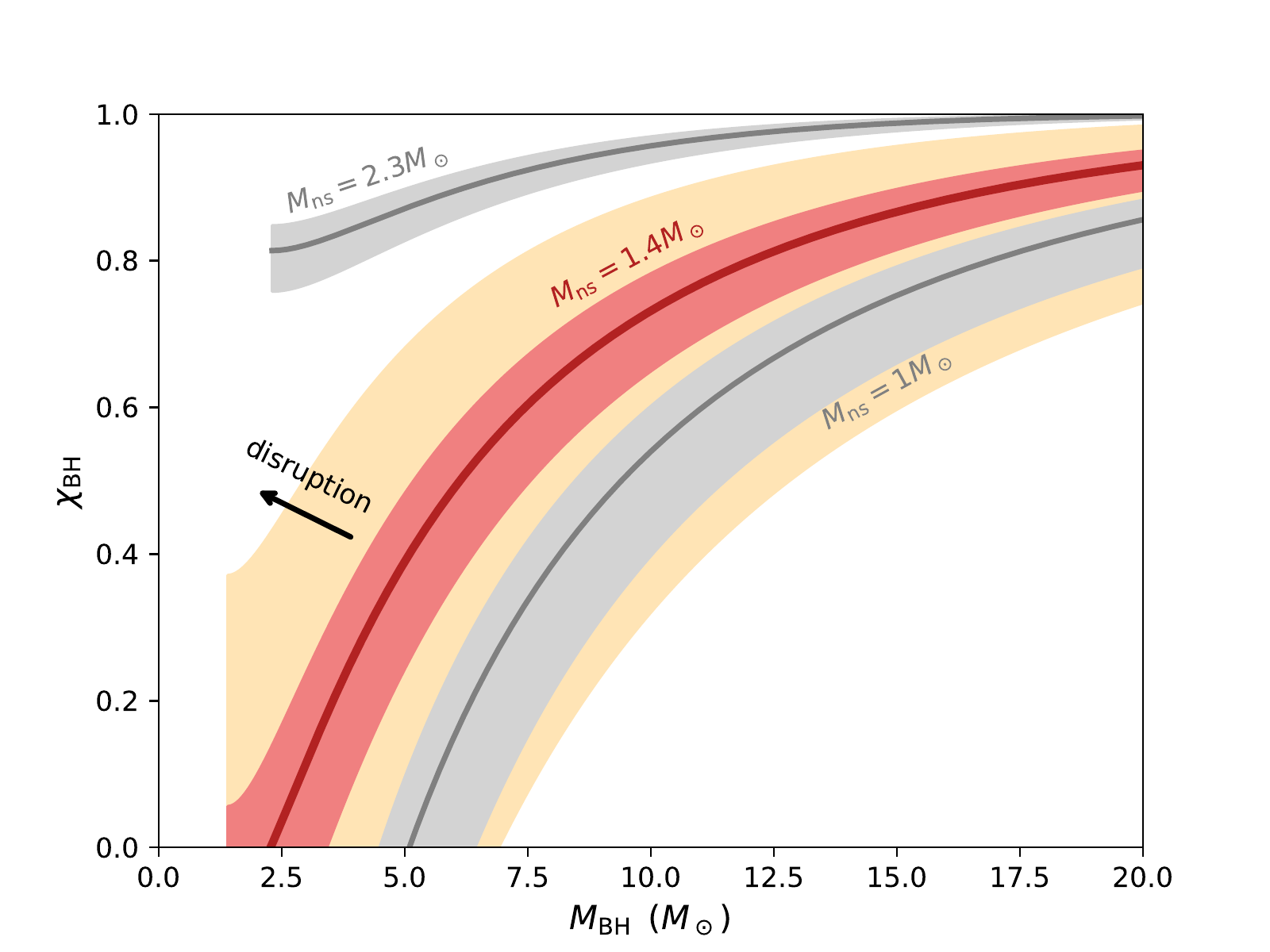}
\caption{Parameter space of neutron-star--black-hole mergers delineating regions where the neutron star is tidally disrupted before merger (upper-left), from those in which the merger occurs without any mass ejection (lower-right). In the latter case, neither a gamma-ray burst nor a kilonova electromagnetic counterpart would be expected. Each curve shows the minimal black hole spin $\chi_{\rm BH}$ required to disrupt a neutron star of a given mass (labeled) and as a function of the black hole mass $M_{\rm BH}$, calculated following \cite{Foucart+18}.
The criterion depends sensitively on the neutron-star radius. Our finding of $R_{1.4M_\odot}=11.0^{+0.9}_{-0.6}$ stringently constrains this parameter space and implies a narrow uncertainty width around each curve (shaded red/grey regions). 
For comparison, the $1.4\,\mathrm{M}_\odot$ curves for weakly-constrained neutron-star radii, $9\,{\rm km} < R_{1.4\,\mathrm{M}_\odot} < 15\,{\rm km}$, span the entire yellow-shaded region, providing only weak predictive power.
Our new constraint on $R_{1.4\,\mathrm{M}_\odot}$ implies that typical neutron stars cannot be disrupted by non-spinning black holes,
except possibly for unusually low black-hole mass.
The grey curves show a rough bound on the parameter space of allowed neutron-star masses, where $M_{\rm ns} \leq M_{\rm max} < 2.3 \,\mathrm{M}_\odot$ as described in the text, and the lower limit $M_{\rm ns}>1\,\mathrm{M}_\odot$ is expected in standard astrophysical neutron-star formation scenarios.
\label{fig:NS_disruption_chi}}
\end{figure}

\clearpage

\begin{table}[t]
\centering
\begin{tabular}{|cc||c|c|}
\hline
Observable                          & Analysis stage    & n$_{\rm sat}$ & 2n$_{\rm sat}$  \\
\hline
\hline
\multirow{3}{*}{$R_{1.4}$[km]}      & Prior & 
$12.1 \pm 2.6$ & $10.9\pm 1.4$  \\
                                    & +GW   & $10.5^{+1.8}_{-1.2}$ & $10.5^{+1.3}_{-1.0}$ \\
                                    & +EM   & $11.2^{+1.2}_{-0.8}$ &  $11.0^{+0.9}_{-0.6}$ \\
\hline
\multirow{3}{*}{$\tilde{\Lambda}$}  & Prior & $330^{+1780}_{-300}$ & $160^{+630}_{-130}$  \\
                                    & +GW   & $180^{+340}_{-100}$  & $190^{+210}_{-100}$  \\
                                    & +EM   & $270^{+260}_{-100}$ & $256^{+139}_{-75}$   \\
\hline
\multirow{3}{*}{$M_{\rm max}$ [$M_\odot$]}  & Prior     & $2.39^{+1.09}_{-0.48}$ & $2.12^{+0.41}_{-0.21}$  \\
                                    & +GW   & $2.01^{+0.33}_{-0.10}$ & $2.01^{+0.34}_{-0.11}$  \\
                                    & +EM   & $2.07^{+0.20}_{-0.14}$ & $2.10^{+0.18}_{-0.17}$  \\
\hline
\multirow{3}{*}{$P_{\rm max}$[MeV/fm$^3$]}  & Prior & $517^{+512}_{-371}$ & $644^{+437}_{-394}$  \\
                                    & +GW   & $730^{+350}_{-380}$ & $730^{+350}_{-440}$  \\
                                    & +EM   & $600^{+380}_{-330}$ & $570^{+320}_{-320}$  \\
\hline
\multirow{3}{*}{$P_{4n_{\rm sat}}$[MeV/fm$^3$]} & Prior & $170^{+182}_{-111}$ & $158^{+142}_{-101}$  \\
                                    & +GW   & $123^{+107}_{-70}$ & $125^{+118}_{-68}$  \\
                                    & +EM   & $154^{+58}_{-49}$  & $161^{+58}_{-46}$  \\
\hline
\end{tabular}

\caption{Summary of the radius of a $1.4 M_{\odot}$ neutron star $R_{1.4\,\mathrm{M}_\odot}$, the tidal polarizability $\tilde{\Lambda}$, the maximum neutron-star mass $M_{\rm max}$, the maximum pressure explored in neutron stars $P_{\rm max}$, and the pressure at four times nuclear saturation density $P_{4n_{\rm sat}}$ at different stages in our analysis. We quote the prior values, values after applying gravitational-wave (GW) constraints, and finally values when both constraints from electromagnetic (EM) observations are applied. Quoted values are the median plus/minus 95th and 5th percentiles.
}
\end{table}

\clearpage

\renewcommand{\figurename}{Supplementary Figure}
\renewcommand\thefigure{\arabic{figure}}
\setcounter{figure}{0} 
 
\section*{Supplementary Material}

\subsection*{Comparison to previous multimessenger constraints}

Recently, Radice and Dai~\cite{Radice&Dai18} and Coughlin et al. \cite{Coughlin+18} also performed multimessenger parameter estimation for GW170817.
Radice and Dai~\cite{Radice&Dai18} imposed a lower limit on the tidal polarizability of the merging neutron stars based on arguments that the mass surrounding the remnant after merger (and that can subsequently become unbound and contribute to the kilonova through secular disk winds~\cite{Metzger+08,Fernandez&Metzger13,Siegel&Metzger17,Fernandez+19}) is strongly correlated with $\tilde{\Lambda}$~\cite{Radice+18}. Combining this lower limit with the GW data and translating their resulting constraints on $\tilde{\Lambda}$ into $R_{1.4\,\mathrm{M}_\odot}$ using an equation-of-state insensitive relation~\cite{De2018}, they find $R=12.2^{+1.0}_{-0.8} \pm 0.2 \, {\rm km}$.
Coughlin et al.~\cite{Coughlin+18} performed joint electromagnetic and GW parameter estimation, directly fitting the kilonova photometry using results from radiative transfer calculations~\cite{Kasen+17} and numerical relativity simulations, along with modeling of the associated gamma-ray burst GRB170817A. Sampling in $\tilde{\Lambda}$ space and similarly translating their results into radius constraints using the equation-of-state insensitive relation of De et al.~\cite{De2018}, they obtained $R \in (11.1,13.4) \pm 0.2 \, {\rm km}$ \cite{Coughlin+18}.

These earlier studies differ from our present work in several respects. First and foremost, we combine systematic nuclear theory with the gravitational wave and electromagnetic observations. We use theoretically motivated equations of state that comply with low-density nuclear experimental data and neutron-star mass measurements while allowing for the most general behavior at large densities, and are therefore able to directly constrain the neutron-star radius rather than the tidal polarizability. This alleviates the need of assuming a universal relation between $\tilde{\Lambda}$ and $R_{1.4\,\mathrm{M}_\odot}$, and allows us to explore the parameter space in the most self-consistent way.
Furthermore, we have adopted conservative assumptions regarding the electromagnetic constraints, based on purely qualitative features of the electromagnetic counterparts.
While quantitatively fitting these counterparts as in Coughlin et al.~\cite{Coughlin+18} provides a promising avenue for future investigation, the uncertainties associated with such modeling are still poorly constrained.
In this respect, it is unsurprising that the lower limit on $R_{1.4\,\mathrm{M}_\odot}$ we obtain here (and which is governed primarily by the electromagnetic constraint) is lower than in these previous studies as we have made conservative assumptions.
As a final note, we point out
that our results are also consistent with early suggestions for a small neutron-star radius ($\leq 12$km) as a possible way to explain the `blue' kilonova component of GW170817 \cite{Nicholl+17}.

\clearpage

\begin{figure}[ht]
\centering
\includegraphics[width=0.8\textwidth]{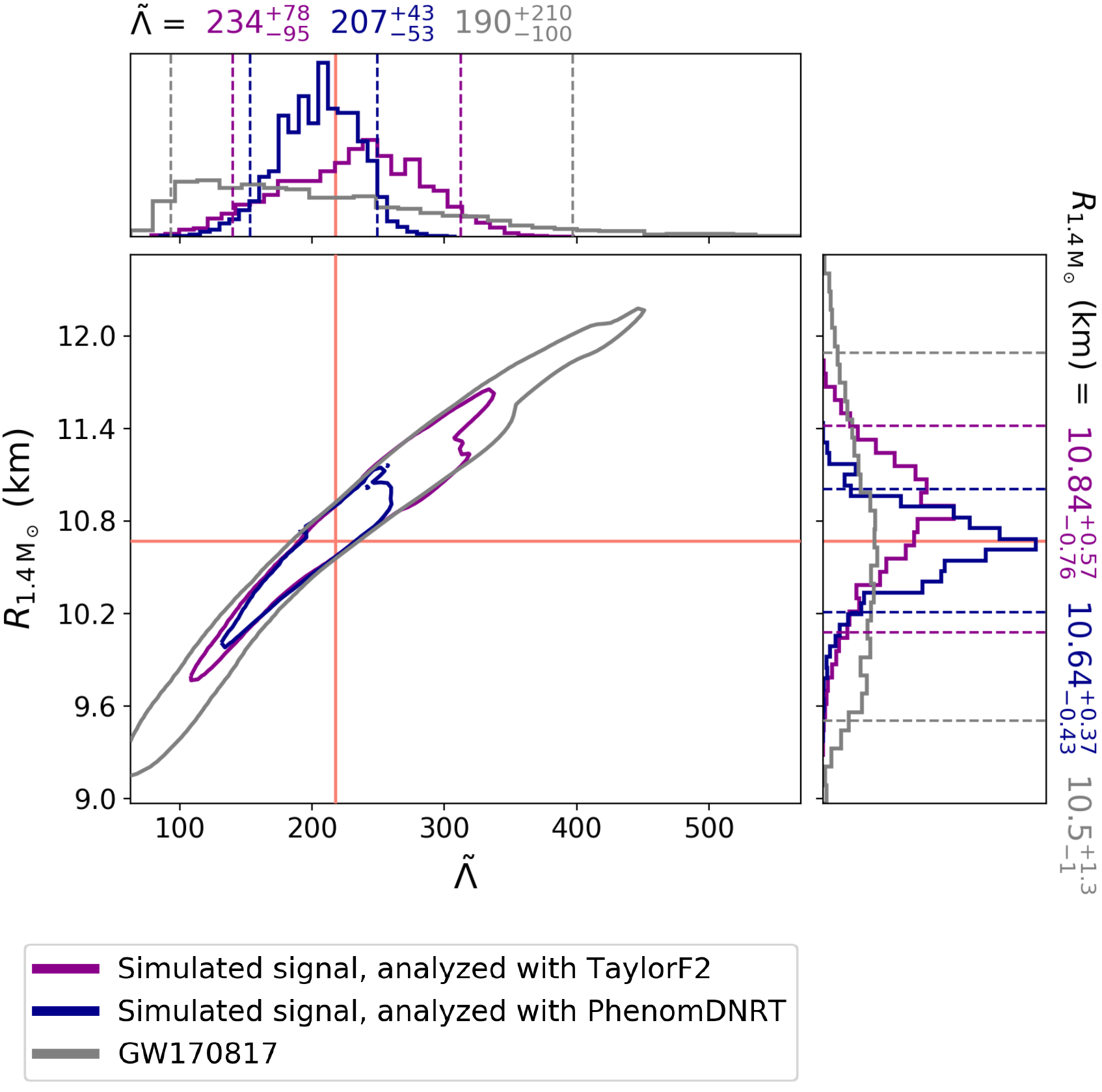}
\caption{Posterior probability distribution of $R_{1.4\,\mathrm{M}_\odot}$ and $\tilde{\Lambda}$ for a simulated GW170817-like signal at a signal-to-noise ratio of $\sim100$, compared to our gravitational-wave only analysis of GW170817. Solid red lines show the parameters of the simulated signal. The top and side panels show 1D marginal posterior distributions for the parameters, with the dashed lines showing the 5th and 95th percentiles. The titles quote the median $\pm$ the 95th and 5th percentile for each parameter. The contours show the 90\% credible region of the 2D marginalized posterior. All of these analyses used the $2n_{\mathrm{sat}}$ prior for the equations of state.
\label{fig:lam_radii_inj_compare}}
\end{figure}

\end{document}